\documentclass[prb,aps,twocolumn,amsmath,amssymb,floatfix,superscriptaddress]{revtex4}

\usepackage{amsmath}
\usepackage{amsfonts}
\usepackage{amssymb}
\usepackage{alphabeta}

\usepackage[dvips]{graphics}

\usepackage{color}
\definecolor{dred}{rgb}{0.75,0,0}
\usepackage{soul}
\usepackage{paralist}
\usepackage[colorlinks=true, citecolor=blue, urlcolor=blue ]{hyperref}

\date{\today}

\begin{document}
	
\title{Bias driven circular current in a ring nanojunction: Critical role of environmental interaction} 

\author{Moumita Mondal}

\affiliation{Physics and Applied Mathematics Unit, Indian Statistical
Institute, 203 Barrackpore Trunk Road, Kolkata-700 108, India}

\author{Santanu K. Maiti}

\email{santanu.maiti@isical.ac.in}

\affiliation{Physics and Applied Mathematics Unit, Indian Statistical
Institute, 203 Barrackpore Trunk Road, Kolkata-700 108, India}

\begin{abstract}
	
The specific role of environmental interaction on bias driven circular current in a ring nanojunction is explored within a tight-binding 
framework based on wave-guide theory. The environmental interaction is implemented through disorder in backbone sites where these sites 
are directly coupled to parent lattice sites of the ring via single bonds. In absence of backbone disorder circular current becomes zero 
for a lengthwise symmetric nanojunction, while it increases with disorder which is quite unusual, and after reaching a maximum it eventually 
drops to zero in the limit of high disorder. The effects of ring-electrode interface configuration, ring-backbone coupling, different types 
of backbone disorder and system temperature are critically investigated. All the studied results are valid for a broad range of physical 
parameters, giving us confidence that the outcomes of this theoretical work can be verified experimentally. To make this 
study self-contained, we also discuss the feasibility of detecting bias-driven circular current and provide design guidelines for 
implementing our proposed quantum system in a laboratory.

\end{abstract}

\maketitle

\section{Introduction}

Nano rings have long been the focus of intense research, revealing numerous fascinating phenomena compared to linear counterparts. 
When a loop conductor interfaces with external electronic baths, it generates a net circular current under specific 
conditions~\cite{cr1,cr2,cr3,cr4,cr5}. While researchers are familiar with transport or junction currents~\cite{jn1,jn2,jn3,jn4}, 
bias-driven circular currents represent a relatively novel phenomenon 
that has yet to be thoroughly investigated. Some groups~\cite{cr1,cr2,cr3,cr4,cr5} have made attempts in this direction previously, but 
it was the pioneering work of Nitzan and his team~\cite{nt1,nt2} around a decade ago that brought it to prominence. Circular currents 
involve current distributions within individual bonds, leading to the emergence of crucial features such as the conducting properties 
of multi-arm loop systems, the specific role of disorder, and the nature of total current flow.

\begin{figure}[ht]
\centering \resizebox*{8.25cm}{4.5cm}{\includegraphics{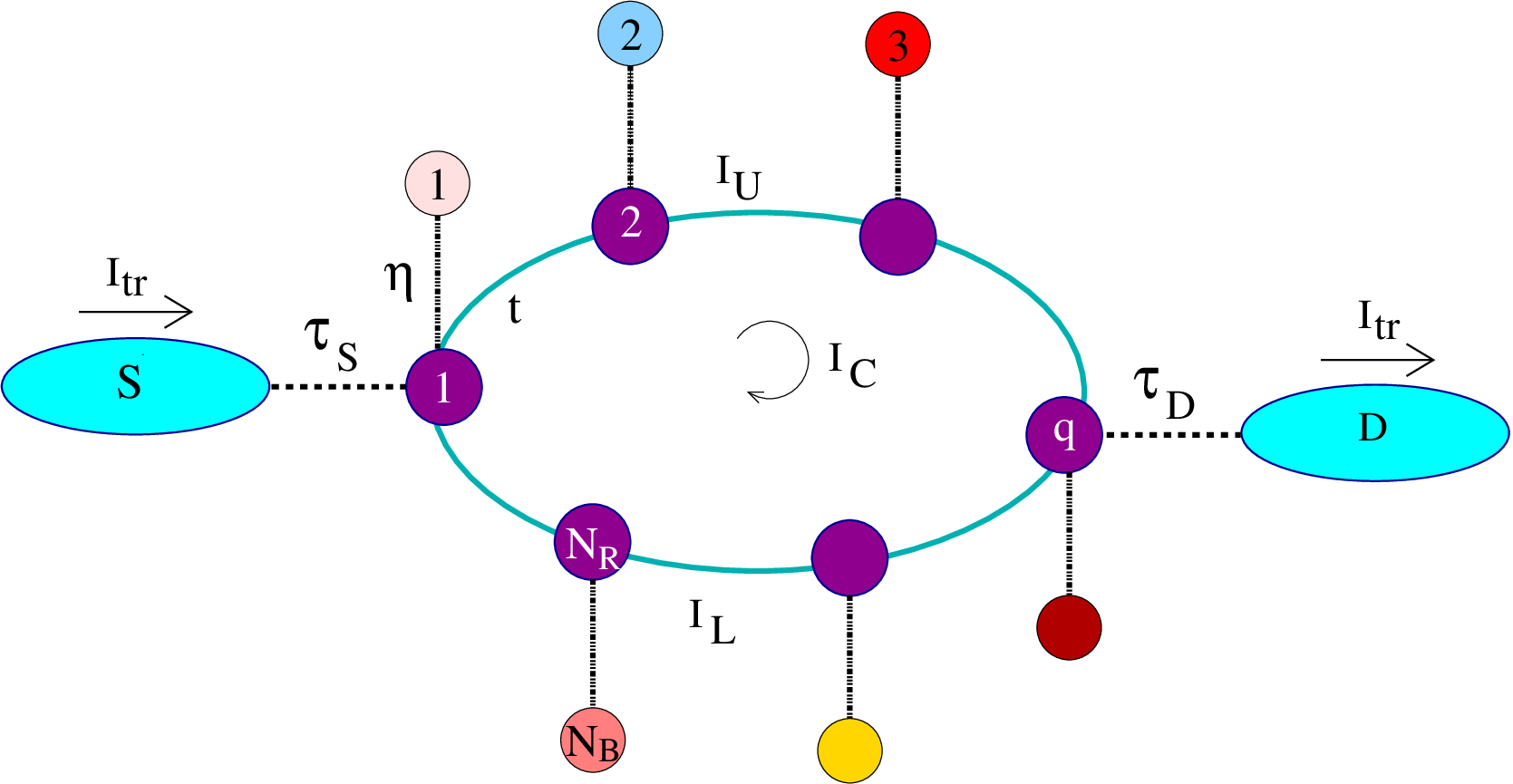}}
\caption{(Color online). Schematic diagram of the junction setup where a nano ring is attached to source and drain electrodes. Each site 
of the ring is connected to a backbone site via a single bond. In presence of finite bias between source (S) and drain (D), a net 
circular current, $I_c$, appears in the ring. $I_{\mbox{\small tr}}$ represents the transport current, commonly referred 
to as the junction current. Disorder is introduced at the backbone sites while keeping the ring sites clean.}
\label{fig1}
\end{figure}

The majority of literature articles have primarily explored the phenomenon of circular currents in nanojunctions under voltage bias, 
typically assuming disorder-free conductors~\cite{cr1,cr2,cr3,cr4,cr5,nt1,nt2,skm1,skm2,rai1}. Little attention has been paid to 
discussing the influence of disorder~\cite{wg1,mm}. {\em However, the impact of environmental interactions, which are often unavoidable in 
experimental settings, has not been well investigated.}
Our current study aims to fill this gap. To address this, we examine a nano ring positioned between source and drain electrodes, with 
each ring site directly linked to a backbone site (refer to Fig.~\ref{fig1}). The environmental interaction is introduced 
phenomenologically by incorporating impurities into these backbone sites~\cite{bb1,bb2,bb3,bb4,bb5,bb6,bb7}.
This protocol is a standard method for assessing how the environment interacts with the system, and it has been utilized in numerous 
studies concerning transport phenomena. Various types of disorder can be considered. It can either be fully uncorrelated 
(random)~\cite{r1,r2,r3,r4,het1}, which is commonly employed, or correlated 
(non-random)~\cite{r0,r5,r6,r7,r8,r9,r10,r11,r12,r13,cn1,cn2,cn3,cn4,cn5,cn6,cn7,cn8}. 
Random disorder is relatively straightforward but requires extensive 
configuration averaging across numerous distinct configurations. Conversely, with correlated disorder, configuration averaging is 
unnecessary, and many nontrivial signatures emerge. One captivating example of correlated disorder is the Aubry-Andr\'{e}-Harper (AAH) 
model~\cite{r5,r6,r7,r8,r9,r10}, renowned for its diverse and intriguing characteristics, which has been extensively explored in 
various contexts.

In this study, we incorporate disorder in backbone sites following the AAH form and explicitly 
discuss its effect on bias-driven 
circular current. Using a tight-binding (TB) framework to illustrate the nanojunction, we obtain the circular current based on the 
well-known wave-guide (WG) theory~\cite{skm2,wg1}. Interestingly, we observe that the `bias driven circular current' increases with 
disorder strength. After reaching a maximum, the current decreases and practically drops to zero in the limit of high disorder strength. 
Thus, backbone disorder, representing environmental interaction, effectively enhances the circular current within a specific region, 
which is quite important in the context of circular current in a loop nanojunction.
Here, it is relevant to note that disorder-assisted transport phenomena have been explored~\cite{da1,da2} in different other quantum 
systems, revealing the fact that disorder can enhance transport current. Considering a tight-binding one-dimensional chain with random 
side sites, Xiong has shown~\cite{da3} with simple mathematical analysis and clear physical arguments that conduction can be improved 
with increasing the disorder strength. This argument matches well with the analysis done by Zhong {\em et al.}~\cite{da4} in their 
work where they have considered shell-doped nanowires in which dopant atoms are distributed in some regions, leaving other parts undoped. 
As the doping concentration increases, conduction improves, unlike to fully disordered systems. Several atypical signatures are expected 
when a perfect conductor is coupled to a disordered region, differing from those seen in a fully disordered system.
To make our present study comprehensive, along with disorder, we also critically investigate the effects of (i) ring-electrode junction 
configuration, which plays a dominant role in transport behavior, (ii) ring-backbone coupling, (iii) other types of disorder, and 
(iv) system temperature. Our detailed numerical calculations reveal that environmental interaction has a strong effect on circular current.

We organize our work as follows. After the brief introduction above, in Sec. II we present the ring nanojunction, its tight-binding 
Hamiltonian, and the necessary theoretical steps for obtaining the results. In a separate appendix some additional mathematical 
steps are given for better understanding. In Sec. III, we critically explain all the findings with 
appropriate physical arguments. In addition, we explore the possibilities of detecting bias driven circular current and 
provide design prescriptions for implementing our proposed quantum system in a laboratory. Finally, we conclude in Sec. IV.

\section{Junction setup, Hamiltonian and Theoretical framework}

\subsection{Nanojunction and the TB Hamiltonian}

Let us start with the junction setup shown in Fig.~\ref{fig1}. A one-dimensional nano ring, possessing $N_R$ lattice sites, 
is clamped between two 
electron reservoirs, commonly referred to as source (S) and drain (D). Each lattice site of the ring (represented by a filled
purple circle) is connected to a backbone site via a single bond. To denote backbone sites are disordered, we use different colors for
those sites. The total number of backbone sites is mentioned by the parameter $N_B$, and in our setup $N_B$ is always identical to
$N_R$. Once a bias is applied across the electrodes, a transport current ($I_{\mbox{\small tr}}$), more commonly referred to as junction 
current, is generated. This current splits into two arms of the ring and reunite at the drain end. Along with the transport current, a
net circular current, specified by $I_c$, can be induced in the ring, under suitable conditions. While the behavior of the transport 
current is well understood, the characteristic features of bias-driven circular current remain relatively unexplored. This study aims 
to investigate these aspects.   

The ring-electrode junction system is simulated within a tight-binding framework. Since the complete system contains different parts,
it is convenient to write the full Hamiltonian, $H$, as a sum of different sub-Hamiltonians related to different parts of the junction and 
the associated coupling, viz, ring, source, drain and the coupling between ring and electrodes~\cite{skm2,wg1}.
The general form of TB Hamiltonian of any part within nearest-neighbor hopping (NNH) approximation looks like
\begin{equation}
H_{\beta}=\sum_n \epsilon_{\beta,n}c_{\beta,n}^{\dagger}c_{\beta,n} + \sum_n t_{\beta} \left(c_{\beta,n}^{\dagger}c_{\beta,n+1}+h.c.\right)
\label{ham}
\end{equation}
where $\beta=S$, $D$, ring (R) and backbone (B). $\epsilon_{\beta,n}$ corresponds to the on-site energy of an electron at site $n$ of the part
$\beta$ and $t_\beta$ is the NNH strength. Now we explicitly mention these TB parameters associated to different
parts of the junction. For the side-attached electrodes we set $\epsilon_{\beta,n}=\epsilon_0$ and $t_\beta=t_0$. The electrodes are 
assumed to be one-dimensional, reflection-less and perfect. The source and drain are coupled to the ring via the coupling parameters 
$\tau_S$ and $\tau_D$, respectively. In the junction setup, it is considered that the source is always connected to site number $1$ of 
the ring, while the drain position (site number $q$) can vary (see Fig.~\ref{fig1}). The physical system sandwiched between source and 
drain contains two different kinds of sites: sites associated to the ring (those are refereed to parent lattice sites) and the backbone
sites. Any ring site is labeled as $\epsilon_{R,n}$ and for the backbone site it is $\epsilon_{B,n}$. In our setup the ring sites are 
clean (viz, all the ring sites are identical), and disorder is introduced only in the backbone sites. Unless specified, we choose 
$\epsilon_{B,n}$ in the form of AAH model~\cite{r7,r8,r9,r10}
\begin{equation}
\epsilon_{B,n}=W \cos(2 \pi b n)
\label{se}
\end{equation}  
where $W$ is the disorder strength and $b$ is an irrational number. Here, the site index $n$ runs over all the backbone 
sites, i.e., from $1$ to $N_B$. As pointed out, another type of disorder can also be taken into 
account. Each backbone site is connected to a parent lattice site via a single bond with the hopping strength $\eta$. For the case of 
rational $b$, the backbone sites become perfect and the periodicity depends on the choice of $b$ and $N_B$. The NNH strength 
in the ring is mentioned by the parameter $t$.

\subsection{Theoretical Prescription}

To understand the description of circular current, let us start with arm currents associated with different arms of the ring geometry. For our 
chosen setup, if the upper and lower arms having the lengths $L_U$ and $L_L$ carry the currents $I_U$ and $I_L$ respectively then the circular
current,  $I_c$, is defined as~\cite{nt1,nt2,skm1}
\begin{equation}
I_c=\frac{I_UL_U + I_LL_L} {L_U + L_L}
\label{ic}
\end{equation}
where $L$ (circumference of the ring) $= L_U + L_L$. When the lengths and status of the two arms are exactly identical it is simple to 
follow that $I_U =-I_L$ and then $I_c$ becomes zero. So, in order to have a non zero $I_c$ we need to break the symmetry between the two 
arms. The symmetry breaking can be done in three ways: (i) by changing the arm lengths or (ii) by setting the status of the two arms 
different when their lengths are identical or (iii) by both. In our case, we set all those conditions one by one.

In TB framework the current in any arm or any single/multiple bonds can be calculated by using the WG theory 
(a standard protocol)~\cite{skm2,wg1}. 
For that first we need to calculate the bond current densities in individual bonds. The current density in any bond of the ring is 
defined as~\cite{nt1,nt2,skm1,skm2}
\begin{equation}
J_{n,n+1}(E)=\left(\frac{2e} {\hbar}\right) \mbox{Im} \left[t C_{R,n}^* C_{R,n+1}\right] 
\label{jc}
\end{equation}
where t is the NNH strength in the ring (mentioned earlier), $e$ and $\hbar$ $(=h/2\pi)$ carry their usual meanings; $C_{R,n}$'s are 
the wave amplitudes, and they are evaluated by solving ($N_R+N_B+2$) coupled equations (for more details, we suggest to see 
Refs.~\cite{skm2,wg1}). The general form of the coupled equations look like
\begin{equation}
(E-\epsilon_{\beta,n}) C_{\beta,n} = \sum_{m} t_{\beta,m} C_{\beta,m}
\label{le}
\end{equation}
where $m$ is the site index referring to the nearest-neighbor sites of $n$. The ($N_R+N_B$) equations come from the ring and backbone 
sites, and other two equations arise from the sites where the ring is coupled to S and D. For a complete understanding 
of these coupled equations, refer to Appendix~\ref{app}, where all equations are explicitly written for a ring nanojunction. Once the 
coefficients, $C_{\beta,n}$, associated with the lattice sites are obtained, we can determine all the bond current densities. By 
summing over the relevant bonds, we can compute the current densities associated with upper and lower arms, denoted as $J_U$ and $J_L$,
respectively.

Integrating current densities over a suitable energy window we find the currents, as a function of voltage, associated with different 
arms. The arm current is expressed as~\cite{nt1,nt2,skm1,skm2}
\begin{equation}
I_{U/L}=\int J_{U/L}(E) (f_S-f_D)\,dE
\label{iu}
\end{equation}
where $f_{S(D)}$ is the Fermi-Dirac distribution function for S(D), and it is 
\begin{equation}
f_{S(D)} = \frac{1} {1+e^{\frac{\left(E-\mu_{S(D)}\right)} {K_B T}}}.
\label{fe}
\end{equation}
Here $T$ is the equilibrium temperature, $K_B$ is the Boltzmann constant, and $\mu_S$ and $\mu_D$ are the electro-chemical potentials of 
S and D respectively. In terms of the equilibrium Fermi energy $E_F$ and bias voltage $V$, $\mu_S$ and $\mu_D$ are written as: 
$\mu_S = E_F + eV/2$ and $\mu_D = E_F - eV/2$.
Once $I_U$ and $I_L$ are found out, the circular current $I_c$ is obtained from the above mentioned definition (Eq.~\ref{ic}).

\section{Results and discussion}

The central focus of this work is to investigate the specific role of backbone disorder on bias driven circular current in a nano ring. 
Unless mentioned, the disorder is introduced in backbone sites in the cosine form following the AAH model. The incommensurate factor `$b$' 
in the expression of $\epsilon_{B,n}$ is chosen as $(1+\sqrt{5})/2$ (golden mean) which is quite common for the AAH model and has been 
used extensively in the literature~\cite{r10,ros}, though any other irrational number can be taken into consideration. At the end of 
this section, the 
effects of other two types of correlated disorders on $I_c$ are also discussed to check the sensitivity of $I_c$ on the nature of the 
disorder. The results are mostly discussed for the lengthwise symmetric ring nanojunction, setting the temperature at zero Kelvin. The 
effects of ring-electrode junction configuration, ring-backbone coupling and temperature are studied in appropriate parts for the sake 
of completion.

For computing the results, the values of some of the physical parameters are kept constant throughout the work, 
\begin{figure}[ht]
\centering \resizebox*{8cm}{10cm}{\includegraphics{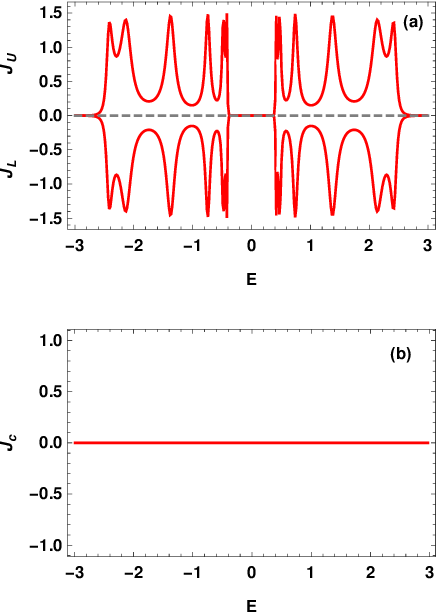}}
\caption{(Color online). Variation of (a) $J_U$ and $J_L$ and (b) total current density $J_c$ ($=J_U+J_L$) as a function of energy 
$E$ for a $10$-site ring in the absence of any disorder ($W=0$) on backbone sites. The upper and lower arms are identical in length.}
\label{fig2}
\end{figure}
and here it is relevant 
to mention them. If not indicated otherwise, we set $\epsilon_0=0$, $t_0=3$, $\tau_S=\tau_D=1$, $\epsilon_{R,n}=0$, $t=1$, $\eta=1$, 
$E_F=0$, $T=0$ and $N_R=10$. The values of other TB parameters that are not fixed, are specified as needed during the discussion. 
All energies are measured in units of eV, and currents are determined in units of mA.

Let us now begin our discussion and analyze the results step by step. To understand the impact of backbone disorder it is indeed meaningful
to start with the setup where all the backbone sites are free from any disorder, which is obtained by setting $W=0$, and to check the 
dependence of current densities associated with the upper and lower arms of the ring, as well as the circular current density. The results
are presented in Fig.~\ref{fig2} where the characteristic behaviors of $J_U$ and $J_L$ are shown in (a), and in (b), the variation of 
the circular current density $J_c$ is depicted. $J_c$ is obtained by summing $J_U$ and $J_L$. Several important features are evident. 
Both for $J_U$ and $J_L$ we have finite peaks around some energies and for other energies they are vanishingly small. For a small window 
across $E=0$, the current densities vanish completely. 
\begin{figure}[ht]
	\centering \resizebox*{8cm}{4.5cm}{\includegraphics{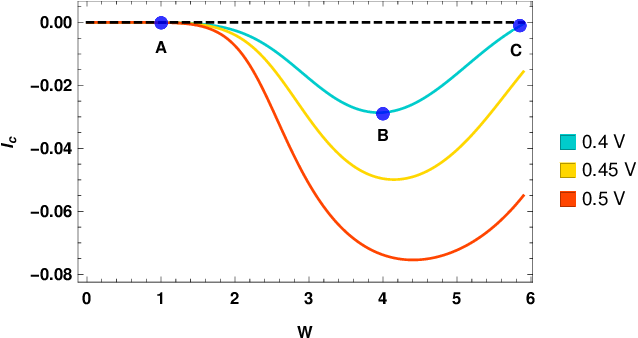}}
	\caption{(Color online). Circular current $I_c$ as a function of backbone disorder strength $W$ for three distinct voltages, denoted 
		by three different colored curves, for a $10$-site ring. Three dot points, A, B and C, on the cyan curve are highlighted to specify the 
		three different values of $W$ (low, moderate and high), where the nature of $I_c$ is critically investigated.}
	\label{fig3}
\end{figure}
These peaks in the current density profiles are associated with the energy eigenvalues of the ring-backbone system clamped
between the electrodes. As the upper and lower arms are identical lengthwise as well as status wise, $J_U$ and $J_L$ become identical 
in magnitude and opposite in sign. In our formulation, positive sign is assigned when current flows in the clockwise direction. The 
nature of the density profile depends on many factors associated with the junction setup. For a symmetric junction configuration, the
peaks are uniform around the discrete energy eigenvalues, and their widths are primarily controlled by the ring-electrode coupling 
strengths, $\tau_S$ and $\tau_D$. For the weak-coupling limit ($\tau_{S(D)}<<t$), the widths are narrow, while they get broadened in the
strong-coupling limit which is specified by the condition $\tau_{S(D)}\sim t$. Since this coupling effect is quite well-known in the 
context of electron transmission, here it is not explicitly discussed (interested readers can follow the Refs.~\cite{cp1,cp2}). Based 
on the nature of 
$J_U$ and $J_L$ (Fig.~\ref{fig2}(a)), the behavior of $J_c$ (Fig.~\ref{fig2}(b)) is easily understood. Across the entire energy window the
net circular density becomes zero, and hence, bias driven circular current is not expected for this case i.e., when the upper and lower 
arms are symmetric to each other.  

The circular current can be generated by breaking the symmetry between the two arms, and here it is done by introducing disorder in the 
backbone sites. As stated earlier, disorder is introduced at all backbone sites while keeping the ring sites free from
disorder. In the presence of disorder, the site energies of all backbone sites ($N_B$) vary from one another, resulting in different 
disorder sequences in the two arms. If the sequences are identical in these two arms, applying disorder becomes meaningless, since, in 
that condition, the arms remain symmetric to each other, resulting in a vanishing net current density, just like in the perfect case 
stated above.
\begin{figure}[ht]
\centering \resizebox*{8cm}{13cm}{\includegraphics{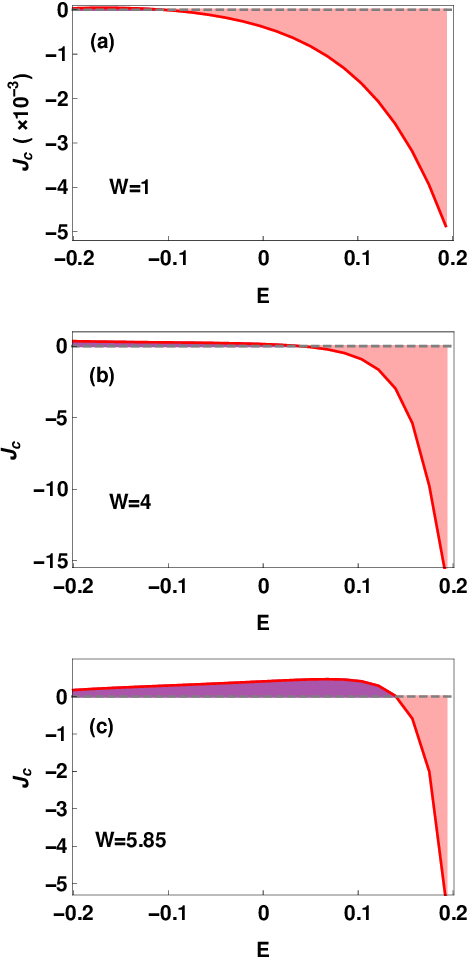}}
\caption{(Color online). Circular current density $J_c$ as a function of energy for three different values of $W$ corresponding to 
points A, B, and C on the cyan curve in Fig.~\ref{fig3}. Color shading is used to highlight the areas under the curves, with two 
different colors indicating the areas above and below the $E=0$ line.}
\label{fig4}
\end{figure}  
Without presenting density profiles, like what are shown for the disorder-free case in Fig.~\ref{fig2}, in 
Fig.~\ref{fig3} we directly present the dependence of circular current, computed at a particular bias voltage, as a function of backbone
disorder strength $W$, since our main concern is to explore the effect of backbone disorder. The results are shown for three distinct 
voltages. Several important features are available.
The overall impression is that the circular current becomes vanishingly small for lower values of $W$, and then it starts increasing with
$W$ and after reaching a maximum it decreases and drops close to zero for too large $W$. This pattern is reflected from all the three curves
computed at three distinct biases. This is one of our central results, and here we explain the nature of the curves for different regimes
of $W$. We choose three points in the cyan curve (other curves can also be taken into account) associated with three different values $W$, 
low, moderate and high, that are represented by A, B and C respectively. For these three cases of $W$, the variations of total circular 
current density $J_c$ are shown in Fig.~\ref{fig4} (exact values of $W$ are mentioned in the sub-figures). The particular energy window
from $-0.2$ to $+0.2$ is used since $V=0.4\,$V and temperature is fixed at zero. When $W=1$, corresponding to A point, the current density 
is extremely small for the entire
energy window (Fig.~\ref{fig4}(a)), and naturally, $I_c$ becomes vanishingly small. On the other hand, for moderate $W$ ($W=4$), $J_c$ is
appreciable and most importantly it is highly asymmetric around $E=0$ (or we can say across $J_c=0$ line) (Fig.~\ref{fig4}(b)). 
The net area of the $J_c$-$E$ curve is thus reasonably large yielding a higher circular current.    
\begin{figure}[ht]
\centering \resizebox*{8cm}{4.5cm}{\includegraphics{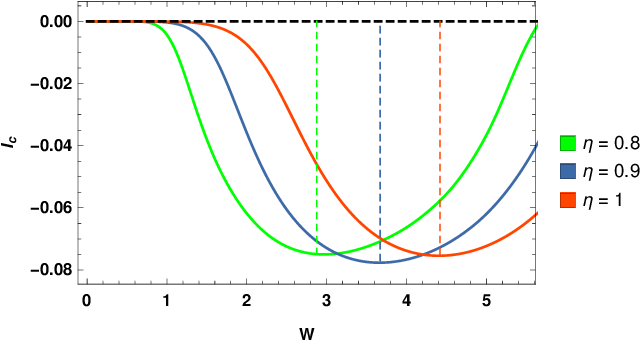}}
\caption{(Color online). Effect of ring-backbone coupling. $I_c$-$W$ curves at three typical values of $\eta$ when the bias voltage is
set at $0.5\,$V. A dashed vertical line is drawn in each curve to clearly locate its extremum point.}
\label{fig5}
\end{figure}
Finally when we reach to the very large value of $W$ (denoting the C point), it is seen from Fig.~\ref{fig4}(c) that the positive and 
negative areas under $J_c$-$E$ curves are quite comparable, resulting in a smaller circular current, like the weak $W$ limit. Thus, form
the behavior of current density profile, the nature of circular current with $W$ can be clearly understood. Now, it is also important to
explain these characteristic features of $I_c$ in different disorder regimes with proper `physical arguments' which are as follows. 
As already pointed out, the primary requirement to have a finite $I_c$      
is to break the symmetry between the upper and lower arms. For $W=0$, all the backbone sites are uniform and the arms become symmetric. 
Once the backbone disorder is introduced, the symmetry is lost, but for weak $W$ the symmetry breaking effect is too small. At the same 
time, within our selected energy window, there are almost no peaks or dips in $J_c$, resulting in a vanishingly small current. With 
increasing $W$, the arms become more asymmetric relative to each other and therefore we get higher circular current with $W$. The situation
becomes counterintuitive beyond a critical disorder where the currents gets reduced with the enhancement of $W$. This can be interpreted 
from the concept of an ordered-disordered scenario~\cite{da4}. The system which is clamped between the electrodes contains two regions. 
In one region
(ring) all the sites are ordered, while for the other region the sites are disordered, and these two regions are coupled to each other.
For weak $W$, the ordered region is affected by the disordered part, and this effect increases with $W$ which is easy to understand. But
the fact is that the coupling between the two regions~\cite{da4} gets weakened with increasing $W$, and for large enough $W$ the ring part is  
almost decoupled from the backbone sites. In that case, the symmetry between the upper and lower arms of the ring is restored,
resulting in a vanishing $I_c$.

Since the nature of circular current is greatly influenced by the interaction between the ordered region i.e., the ring system and 
the disordered backbone region, here it is important to check the effect of $\eta$ on $I_c$. In Fig.~\ref{fig5} we plot $I_c$ as 
a function of $W$ at three typical values of $\eta$ considering the lengthwise symmetric ring nanojunction with $N_R=10$. 
\begin{figure}[ht]
	\centering \resizebox*{8cm}{4.5cm}{\includegraphics{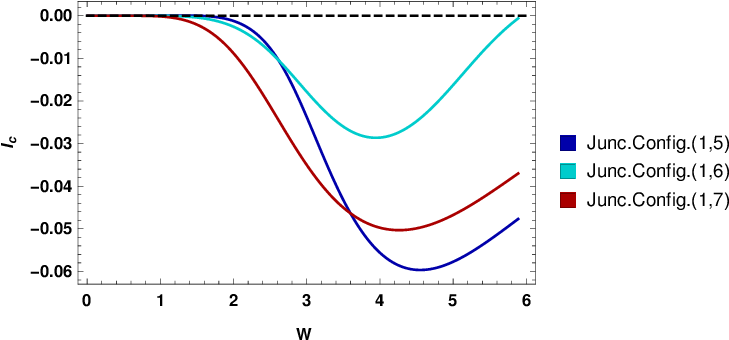}}
	\caption{(Color online). Effect of ring-electrode interface geometry. Variation of $I_c$ as a function of backbone disorder $W$ for 
		three different junction configurations. Along with the length-wise symmetric configuration, two asymmetric configurations are considered 
		where the drain is coupled to sites $5$ and $7$ respectively of a $10$-site ring. Here we set $V=0.4\,$V.}
	\label{fig6}
\end{figure}
All the three curves appear identical in nature. The key observation is that, the critical $W$, where $I_c$ reaches it extremum, shifts 
towards higher $W$ with increasing $\eta$. The underlying mechanism depends on the strength of the connection between the ordered 
and disordered parts. A stronger connection requires a higher $W$ to decouple these two regions, from which point the symmetry between 
the two arms of the ring begins to reemerge.

The atypical behavior of $I_c$ with $W$ which is discussed above perfectly holds for the other ring-electrode junction configurations as 
well, which we claim from the curves shown in Fig.~\ref{fig6}. In this figure, three junction configurations are taken into account, two are 
length-wise asymmetric and one for length-wise symmetric, and the results are worked out for a typical bias voltage $V=0.4\,$V. Along with  
these, some other configurations are also checked and in each case the dependence of $W$ on $I_c$ remains exactly same. It is well-known 
that ring-electrode configuration plays a significant role on transport behavior, due to the modification of quantum interference of 
the wave functions associated with different branches of the ring, and therefore, the maxima points (in the $-$ve side of each curve)
get shifted, but the overall response of $I_c$ with $W$ is unchanged.

To inspect the robustness i.e., whether the above discussed nature is specific to AAH type disorder or it is general with respect to the 
other type of backbone disorder, here we check the behavior of $I_c$-$W$ curve in presence of other types of backbone disorder. We choose 
two different types of disorder in backbone sites: one is the Fibonacci type, and the other is the Bronze Mean (BM) 
one~\cite{r11,r12,r13}. These are quite common examples of correlated disorders and they can be constructed using two different kinds 
of atomic sites, say $A$ and $B$, by arranging them according to specific rules~\cite{r11,r12,r13}, unlike AAH type where all the sites 
are different. The inflation rule for
\begin{figure}[ht]
\centering \resizebox*{8cm}{9cm}{\includegraphics{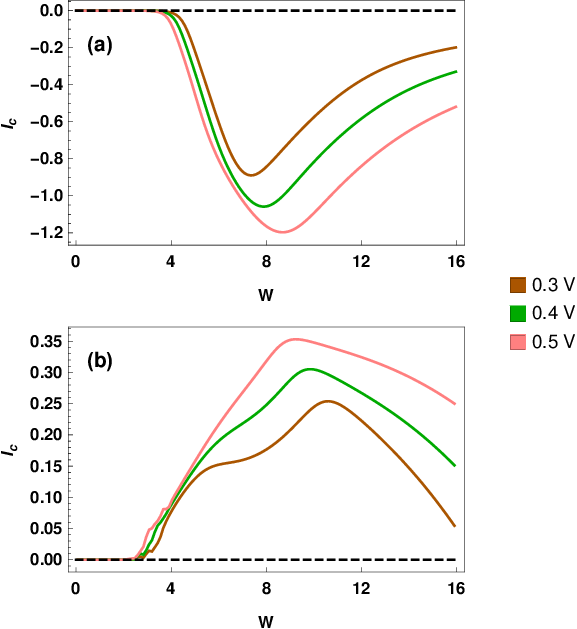}}
\caption{(Color online). $I_c$-$W$ characteristics at three typical bias voltages for the ring system with (a) Fibonacci type backbone 
disorder and (b) BM type backbone disorder. Here we choose $N_R=13$, and connect the source and drain electrodes so that the length 
difference between the upper and lower arms equals the length of a single bond. The other physical parameters are: $t_0=5$, 
$\tau_S=\tau_D=2$, $t=2$, $\eta=2$, and $\epsilon_A=-\epsilon_B=0.5$.}
\label{fig7}
\end{figure}
the Fibonacci sequence is: $A\rightarrow AB$ and $B\rightarrow A$. So the first few Fibonacci generations are $A$, $AB$, $ABA$, $ABAAB$, etc.
On the other hand, the inflation rule for the BM sequence is: $A\rightarrow AAAB$ and $B\rightarrow A$, and here the first few generations are:
$A$, $AAAB$, $AAABAAABAAABA$, etc. For these two different atomic sites ($A$ and $B$) we refer the site energies $\epsilon_A$ and 
$\epsilon_B$ respectively,
and their strengths are specified by the parameter $W$ (like the AAH case). Inserting the Fibonacci and BM disorders in the backbone sites, 
the variations of circular current with respect to $W$ are shown in Fig.~\ref{fig7}. 
Here we consider a $13$-site ring, to have a comparative
ring size with our previously studied AAH case ($10$-site rings cannot be obtained with these sequences). As the ring size is odd, identical 
arm lengths are not possible, and we couple the electrodes in such a way that the length difference between the two arms is just a single
bond. For this particular figure (Fig.~\ref{fig7}), the chosen TB parameters are also quite different compared to the other figures. It is
just because to capture all the information within the varied $W$ window. We make $t_0$ quite large, and thus, we proportionately change
other parameters such that energy band widths of the electrodes are larger than the system placed between them. In each type of backbone
disorder, the variation of $I_c$ with $W$ is shown for three distinct bias voltages, and both for the two types of disorders the nature
of $I_c$ with $W$ remains exactly similar with what is already obtained for the AAH case.
\begin{figure}[ht]
	\centering \resizebox*{8cm}{4.5cm}{\includegraphics{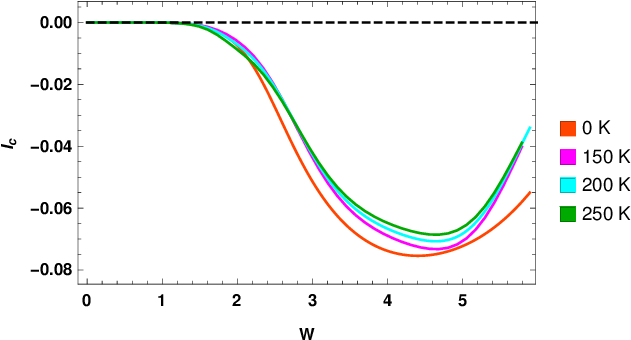}}
	\caption{(Color online). Effect of temperature. Variation of $I_c$ for the ring system possessing AAH type backbone disorder as a 
		function of $W$ at three different temperatures. The result of zero temperature is superimposed for a clear comparison. Here we 
		take $N_R=10$, $\eta=1$ and $V=0.5\,$V. The electrodes are coupled in a lengthwise symmetric configuration.}
	\label{fig8}
\end{figure} 
The only difference is the sign reversal. 
The circular current can have both positive and negative signs (unlike transport current which always exhibits one sign for a 
particular polarity) as it depends on the contributing peaks and dips in the current density profile. The sign of $I_c$ depends on
which dominates among the peaks and dips of $J_c$. Figure~\ref{fig7} clearly suggests that the overall signature of $I_c$ with $W$ remains
unchanged with the type of disorder. Here we would like to note that, in addition to these three distinct types of disorder, the behavior
of $I_c$ remains the same with other different backbone configurations which we firmly confirm through our detailed numerical calculations,
and to avoid any repetition, we do not add those results. 

For our chosen quantum system, it is established that electrical conduction can be monitored in different ways. In particular, backbone disorder plays a crucial role, as it causes the system to map onto an ordered-disordered separated structure, where the coupling between the ordered and disordered regions is significant. Other influential factors include the ring-electrode junction configuration, which directly affects quantum interference, and the ring-electrode coupling. The latter can be selectively tuned in a suitable laboratory setting and has already been examined in various junction setups.

The results analyzed so far are computed for zero temperature. For a more realistic scenario it is also crucial to examine the role of 
temperature. The temperature dependence enters into the current expression through the Fermi-Dirac distribution functions, $f_S$ and $f_D$,
associated with the source and drain electrodes, respectively. The effect of temperature is presented in Fig.~\ref{fig8}, where 
$I_c$-$W$ curves are shown for three distinct temperatures. The result of zero temperature is also superimposed for comparison. Apart from 
a slight reduction of current with temperature, we find that the overall nature of $I_c$-$W$ curve remains unaltered. The reduction of 
current due to temperature can be understood from the following arguments. At absolute zero temperature, the current is obtained by
\begin{figure}[ht]
	\centering \resizebox*{8cm}{5cm}{\includegraphics{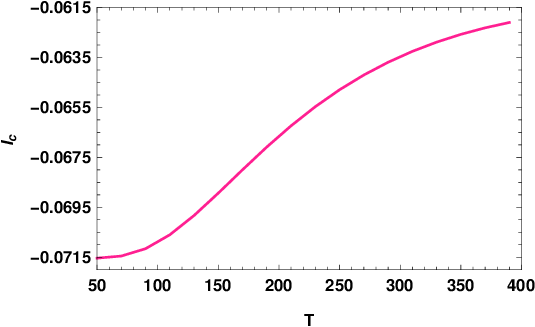}}
	\caption{(Color online). Dependence of $I_c$ with temperature (measured in Kelvin) for the identical setup as mentioned in 
		Fig.~\ref{fig8}. Here $W$ is fixed at $4\,$eV.}
	\label{fig9}
\end{figure}
integrating the current density profile within the energy window from $E_F-eV/2$ to $E_F+eV/2$. Within this energy zone, depending on the
dominating peaks and dips we get the net current. Here the chance of mutual cancellations is relatively less. On the other hand, when 
the temperature is finite, we need to consider the full available energy window where the chance of mutual cancellations may slightly 
increase though it depends on many other factors, especially $W$, and depending on the weight factor ($f_S-f_D$) we get the resultant 
current.  

A more comprehensive dependence of $I_c$ on temperature is given in Fig.~\ref{fig9}, where the variation of $I_c$ at a particular 
voltage is shown by continuously changing the temperature in a broad range. A smooth reduction of $I_c$ with temperature is obtained,
following the above arguments. The key aspect is that, $I_c$ remains finite even at very high temperatures. 

\vskip 0.2cm
\noindent
$\blacksquare$ {\bf Possible ways to detect bias driven circular current}: It is quite significant to explore how bias driven circular current can be verified through experimental observations. Detecting such effects in a practical setting presents both challenges and opportunities. Two possible approaches can be considered~\cite{nt1} for their experimental observation.
The first approach involves examining the spectral response of magnetic ions that are positioned either on or near the quantum ring. 
These ions interact with the local magnetic field induced in their vicinity due to the circular current, and their spectral characteristics 
can reveal valuable information about the underlying magnetic effects. The second approach focuses on the response of local magnetic moment that develops within the ring itself to an external magnetic field. By studying the behavior of magnetic moment, it is possible to infer details of the induced magnetic field generated by the bias driven circular current. Both methods offer promising avenues for experimentally probing this theoretically studied phenomenon, bridging the gap between theoretical predictions and real experimental observations.

\vskip 0.2cm
\noindent
$\blacksquare$ {\bf Possible design prescriptions for the junction}: Here, it is pertinent to briefly describe how our proposed quantum system can be designed in the laboratory. The system consists of two main components: a nano ring and backbone sites. Several novel strategies are available for fabricating a nano ring, including electron-beam lithography~\cite{exp1,exp2}, ion beam milling~\cite{exp3,exp4}, UV lithography~\cite{exp5,exp6}, annealing techniques~\cite{exp7,exp8}, and more. The backbone sites, which can be considered discrete quantum dots (QDs), are positioned such that each ring site is connected to one backbone site via a tunnel junction. Using suitable gate electrodes, the coupling strength between the ring and the backbone sites can be controlled. Additionally, the backbone site energies can be modulated, either randomly or deterministically, through local gate voltage variations or laser-induced potential fluctuations.

\section{Closing Remarks}

In this work, we examine the critical role of environmental interactions on bias-driven circular currents in a nano ring. The 
environmental effects are phenomenologically incorporated by connecting the ring sites to disordered backbone sites, each directly coupled 
to a parent lattice site of the ring via a single bond. The ring is sandwiched between two contact electrodes, source and drain. When a 
finite bias is applied between these electrodes, a net circular current is induced in the ring. The characteristics of this current are 
studied in detail under various input conditions.
 
Using a tight-binding framework to describe the quantum system, all results are derived based on wave-guide theory. The effect of 
backbone disorder is particularly noteworthy. As the disorder strength increases, the current magnitude initially rises, reaches a maximum, 
and then decreases, eventually vanishes at very high disorder strengths. These behaviors are thoroughly analyzed with appropriate 
mathematical results and physical explanations. Additionally, it is established that the effect of disorder remains largely unchanged 
regardless of the type of disorder. Temperature dependence is also discussed considering a more realistic scenario, showing that 
significant current can still be obtained over a wide temperature range. At the end, we briefly discuss the detection 
mechanisms of bias driven circular current and highlight possible strategies for realizing our proposed model in a suitable laboratory.
 
Finally, we emphasize that if environmental interactions is controlled through suitable laboratory methods, transport behavior can be selectively regulated. This potential for control is highly interesting and important. Moreover, controlling the spin-dependent bias-driven circular current through environmental interaction presents another intriguing phenomenon. Various spin-dependent scattering 
factors can be considered~\cite{def1,def2,def3,spi1,spi2}, such as spin-moment interactions in magnetic systems, spin-orbit interactions, Zeeman splitting, etc. In each case, distinct non-trivial signatures can be expected. These aspects will be explored in our forthcoming work.  

\appendix
\section{Coupled equations involving wave amplitudes at different lattice sites}
\label{app}

\begin{figure}[ht]
\centering \resizebox*{8.5cm}{4.5cm}{\includegraphics{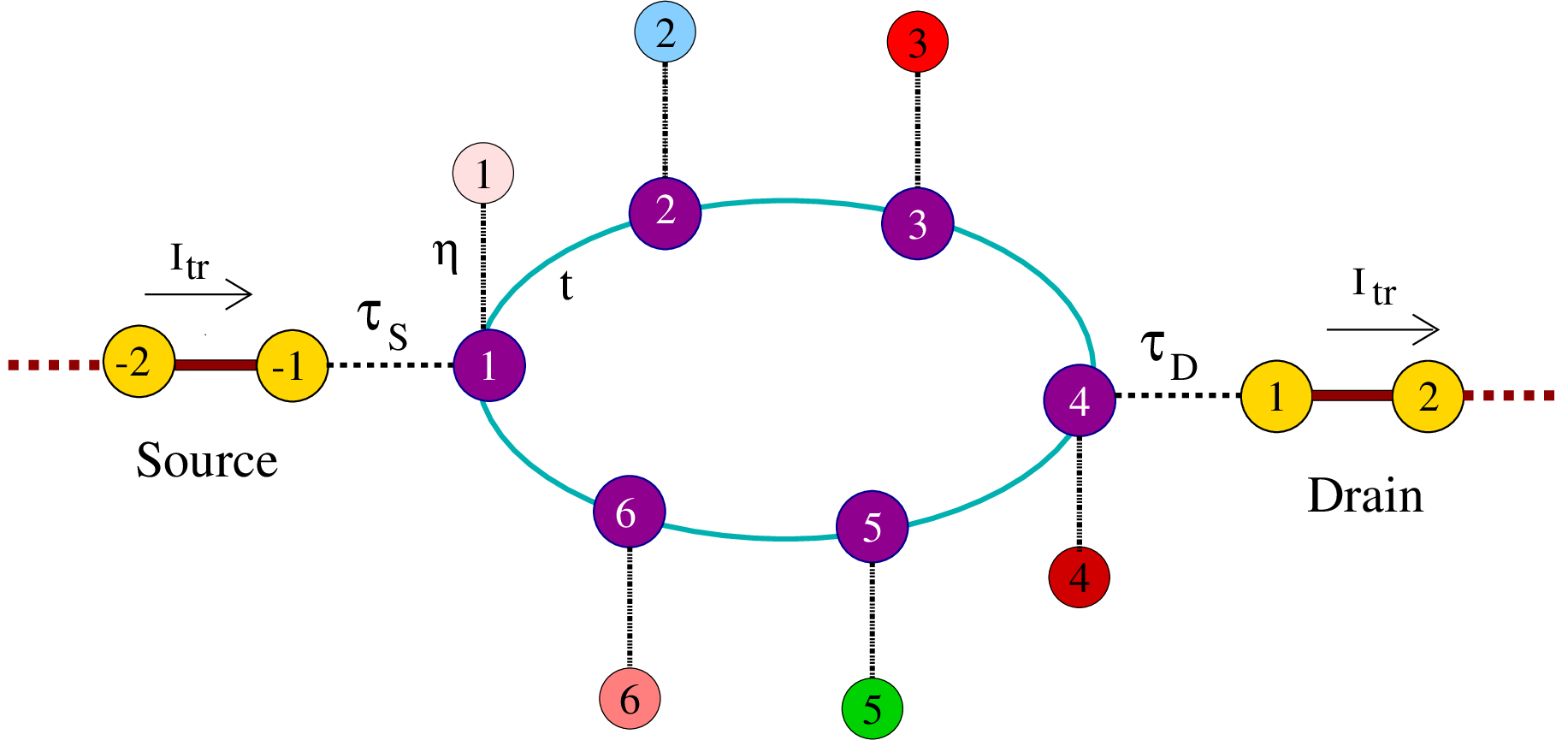}}
\caption{(Color online). Schematic diagram of a ring nanojunction, where a six-site ring is coupled to one-dimensional source and drain electrodes. These electrodes are attached to sites $1$ and $4$ of the ring. Each site of the ring is further connected to a backbone site through a single bond.}
\label{apfig}
\end{figure}

Referring to Fig.~\ref{apfig}, which illustrates a $6$-site ring coupled to source and drain electrodes,
here we express all $14$ equations corresponding to the $14$ lattice sites -- $6$ associated with the ring sites, $6$ with the backbone sites, and the remaining $2$ with the side-attached electrodes.
For the electrodes, these two sites are site number $-1$ of S and site number $1$ of D, as these two sites are directly coupled to the ring sites. The equations are as follows:

\begin{eqnarray}
\left(E-\epsilon_{R,1}\right)C_{R,1}&=&tC_{R,2}+tC_{R,6}+\eta C_{B,1}+\tau_S C_{S,-1},\nonumber\\
\left(E-\epsilon_{R,2}\right)C_{R,2}&=&tC_{R,1}+tC_{R,3}+\eta C_{B,2},\nonumber\\
\left(E-\epsilon_{R,3}\right)C_{R,3}&=&tC_{R,2}+tC_{R,4}+\eta C_{B,3},\nonumber\\
\left(E-\epsilon_{R,4}\right)C_{R,4}&=&tC_{R,3}+tC_{R,5}+\eta C_{B,4}+\tau_D C_{D,1},\nonumber\\
\left(E-\epsilon_{R,5}\right)C_{R,5}&=&tC_{R,4}+tC_{R,6}+\eta C_{B,5},\nonumber\\
\left(E-\epsilon_{R,6}\right)C_{R,6}&=&tC_{R,5}+tC_{R,1}+\eta C_{B,6},\nonumber\\
\left(E-\epsilon_{B,1}\right)C_{B,1}&=&\eta C_{R,1},\nonumber\\
\left(E-\epsilon_{B,2}\right)C_{B,2}&=&\eta C_{R,2},\nonumber\\
\left(E-\epsilon_{B,3}\right)C_{B,3}&=&\eta C_{R,3},\nonumber\\
\left(E-\epsilon_{B,4}\right)C_{B,4}&=&\eta C_{R,4},\nonumber\\
\left(E-\epsilon_{B,5}\right)C_{B,5}&=&\eta C_{R,5},\nonumber\\
\left(E-\epsilon_{B,6}\right)C_{B,6}&=&\eta C_{R,6},\nonumber\\
\left(E-\epsilon_{S,-1}\right)C_{S,-1}&=&\tau_S C_{R,1}+t_0 C_{S,-2},\nonumber\\
\left(E-\epsilon_{D,1}\right)C_{D,1}&=&\tau_D C_{R,4}+t_0 C_{D,2}.
\end{eqnarray}
Assuming a plane wave incidence of unit amplitude from the source end, the wave amplitude at $m$th site of the source electrode 
can be expressed
as $C_{S,m}=e^{ik(m+1)}+re^{-ik(m+1)}$, while for the outgoing wave in the drain electrode, the wave amplitude is given by 
$C_{D,m}=\tau e^{ikm}$. Here $r$ and $\tau$ denote the reflection and transmission amplitudes, respectively, and $k$ is the wave vector
related to the TB parameters through the energy dispersion relation $E= \epsilon_{0}+ 2t_0 \cos(k)$. By solving the above equations, we 
find all the required coefficients as a function of energy $E$.

%\section*{DATA AVAILABILITY STATEMENT}

%All necessary data supporting the findings of this study are included.

%\section*{DECLARATION}

%{\bf Conflict of interest}: The authors declare no conflict of interest.

\end{document}